\documentclass[aps,prd,amsmath, amssymb,eqsecnum,
nofootinbib,showpacs,tightenlines]{revtex4}
\global\arraycolsep=2pt
\usepackage{bm}
\usepackage{epsfig}

\newcommand{\ben}{\begin{equation*}}
\newcommand{\een}{\end{equation*}}
\newcommand{\bean}{\begin{eqnarray*}}
\newcommand{\eean}{\end{eqnarray*}}

\newcommand{\nn}{\nonumber}
\newcommand{\be}{\begin{equation}}
\newcommand{\ee}{\end{equation}}
\newcommand{\bea}{\begin{eqnarray}}
\newcommand{\eea}{\end{eqnarray}}
\begin{document}
\title{From multiple scattering to van der Waals interactions: exact
results for eccentric cylinders}

\author{Kimball A. Milton} 
\email{milton@nhn.ou.edu}
\homepage{http://www.nhn.ou.edu/

\author{Prachi Parashar}
\email{prachi@nhn.ou.edu}

\author{Jef Wagner}
\email{wagner@nhn.ou.edu}
\affiliation{Oklahoma Center for High Energy Physics 
and Homer L. Dodge Department of Physics and Astronomy,
University of Oklahoma, Norman, OK 73019, USA}
\date{\today}

\begin{abstract}
In this paper, dedicated to the career of Iver Brevik, we review
the derivation of the retarded van der Waals
or Casimir-Polder interaction between polarizable
molecules from the general multiple scattering formulation of
Casimir interactions between bodies.  We then apply this van der
Waals potential to examine the interaction between tenuous cylindrical
bodies, including eccentric cylinders, 
and revisit the vanishing self-energy of a tenuous dielectric
cylinder.  In each case, closed-form expressions are obtained.
\end{abstract}

\maketitle

\section{Introduction}

Since the earliest calculations of fluctuation forces between bodies
\cite{Casimir:1948dh},
that is, Casimir or quantum vacuum forces, multiple scattering methods
have been employed.  Rather belatedly, it has been realized that
such methods could be used to obtain accurate numerical results
in many cases \cite{Wirzba:2007bv, Bordag:2008gj, maianeto08, Emig:2007cf}. 
These results allow us to transcend the limitations of the
proximity force theorem (PFT) \cite{pft,derjaguin}, and so make better
comparison with experiment, which typically involve curved
surfaces. (For a review of the experimental
situation, see Ref.~\cite{Onofrio:2006mq}.)

These improvements in technique were inspired in part by the
development of the numerical Monte-Carlo worldline method of 
Gies and Klingm\"uller \cite{Gies:2006cq,Gies:2006bt,Gies:2006xe,Gies:2003cv}
but the difficulty with this latter
method lies in the statistical limitations of Monte Carlo methods 
and in the complexity of incorporating electromagnetic boundary conditions.
Optical path approximations have been studied extensively for many years,
with considerable success 
\cite{Scardicchio:2004fy,Scardicchio:2005di,Fulling03,fulling08}. 
However, there always remain uncertainties 
because of unknown errors in excluding diffractive effects. 
Direct numerical methods \cite{capasso,rodriguez}, 
based on finite-difference
engineering techniques,  may have promise, but the requisite precision
of 3-dimensional calculations may prove challenging \cite{rod}.

The multiple scattering formalism, which is in principle exact,
dates back at least into the 1950s \cite{krein,renne}.  Particularly
noteworthy is the seminal work of Balian and Duplantier \cite{balian}.
(For more complete references see Ref.~\cite{Milton:2007wz}.)
This technique, which has been brought to a high state of perfection
by Emig et al.\ \cite{Emig:2007cf}, 
has concentrated
on numerical results for the Casimir forces between conducting and
dielectric bodies such as spheres and cylinders.  Recently, we have
noticed that the multiple-scattering method can yield exact, closed-form 
results for bodies that are weakly coupled to the quantum field
\cite{Milton:2007gy,Milton:2007wz}. 
This allows an exact assessment
of the range of applicability of the PFT.  The calculations there,
however, as those in recent extensions of our methodology
\cite{CaveroPelaez:2008tj}, 
have been restricted to scalar fields with $\delta$-function potentials,
so-called semitransparent bodies.  
(These are closely related to plasma shell models \cite{Bordag:2008gj,
Bordag:2008rc,Bordag:2007zz,Bordag:2006kx}.) The
technique was recently extended to dielectric bodies \cite{Milton:2008vr},
characterized by a permittivity $\varepsilon$.
Strong coupling would mean a perfect metal, $\varepsilon\to\infty$, while
weak coupling means that $\varepsilon$ is close to unity.

In this paper we will give details of the formalism, and show how
in weak coupling (dilute dielectrics) we recover the sum of
Casimir-Polder or retarded van der Waals forces between atoms.
 Exact results have been found in the past in such
summations, for example for the self-energy of a dilute dielectric
sphere \cite{Milton:1997ky} or a dilute dielectric cylinder
\cite{Nesterenko:1997fq}.  Thus it is not surprising that exact results
for the interaction of different dilute bodies can be obtained.
It is only surprising that such results were not found much earlier.
(We note that the additive approximation has been widely used in the
past, for example, see Ref.~\cite{Bordag:2001qi}, but here the method is
exact.  Also, there are many exact computations for non-retarded London forces
betwen bodies, e.g., Ref.~\cite{Franchek}, but these results can only
apply to very tiny objects on the nanometer scale.) 
In our previous letter \cite{Milton:2008vr} we
considered the force and torque on a slab of finite extent above an 
infinite plane, and the force between spheres and parallel cylinders
outside each other.  Here we will examine further cylindrical
geometries, such as concentric cylinders, and eccentric circular
cylinders with parallel axes, in cases where the
dielectric materials do not overlap.  We will prove that the results can
be obtained by analytic continuation of the energies found earlier
for non-contained bodies.  Finally, we will re-examine the self-energy
of a dielectic cylinder \cite{Nesterenko:1997fq}.

\section{Green's dyadic formalism}

For electromagnetism, we can start from the formalism of Schwinger
\cite{Schwinger:1977pa}, which is based on the electric
Green's dyadic $\bm{\Gamma}$.  This object can be identified
as the one-loop vacuum expectation value of the correlation function
of electric fields,
\be
\bm{\Gamma}(\mathbf{r},t;\mathbf{r'},t')=
i\langle \mbox{T} \{\mathbf{E(r},t)
\mathbf{E(r'},t')\}\rangle.\label{vev}
\ee  Alternatively, we regard the Green's dyadic as the propagator
between a polarization source $\mathbf{P}$ and a phenomenological
field $\mathbf{E}$ (where $x^\mu=(\mathbf{r},t)$):
\be
\mathbf{E}(x)=\int (dx') \bm{\Gamma}(x,x)\cdot \mathbf{P}(x').
\ee
We will only be contemplating static geometries, so it is convenient
to consider a specific frequency $\omega$, as introduced through a Fourier
transform,
\be
\bm{\Gamma}(x,x')=\int\frac{d\omega}{2\pi}e^{-i\omega(t-t')}\bm{\Gamma}
(\mathbf{r,r'};\omega),\label{ftgd}
\ee
in terms of which the Maxwell equations in a region where the permittivity
$\varepsilon(\omega)$ and the permeability $\mu(\omega)$ are constant
in space read
\begin{subequations}
\bea
\bm{\nabla}\times\bm{\Gamma}&=&i\omega\bm{\Phi},\quad \bm{\nabla\cdot\Phi}=0,
\\
\frac1\mu\bm{\nabla}\times\bm{\Phi}&=&-i\omega\varepsilon\bm{\Gamma'},
\quad \bm{\nabla\cdot\Gamma'}=0,
\eea
\end{subequations}
where we have introduced $\bf{\Gamma'}=\bm{\Gamma}+\mathbf{1}/\varepsilon$,
where the unit dyadic includes a spatial $\delta$ function.
The two Green's dyadics given here satisfy the following 
inhomogenous Helmholtz equations,
\begin{subequations}
\bea
(\nabla^2+\omega^2\varepsilon\mu)\bm{\Gamma'}&=&-\frac1\varepsilon
\bm{\nabla}\times(\bm{\nabla}\times\mathbf{1}),\label{gprimeeq}\\
(\nabla^2+\omega^2\varepsilon\mu)\bm{\Phi}&=&i\omega\mu
\bm{\nabla}\times\mathbf{1}.
\eea
\end{subequations}
In the following, it will prove more convenient to use, instead
of Eq.~(\ref{gprimeeq}),
\be
\left(\frac1{\omega^2\mu}\bm{\nabla\times\nabla\times}-\varepsilon\right)
\bm{\Gamma}=\mathbf{1}.\label{geq}
\ee

In the presence of a polarization source, the action is, in
symbolic form,
\be
W=\frac12\int \mathbf{P}\cdot\bm{ \Gamma}\cdot\mathbf{ P},\label{w}
\ee
so if we consider the interaction between bodies characterized by
particular values of $\varepsilon$ and $\mu$, the change in the
action due to moving those bodies is
\be
\delta W=\frac12\int \mathbf{P}\cdot \delta\bm{\Gamma}\cdot \mathbf{P}
=-\frac12\int \mathbf{E}\cdot\delta
\bm{\Gamma}^{-1}\cdot\mathbf{E},\label{dw}
\ee
where the symbolic inverse dyadic, in the sense of Eq.~(\ref{geq}),
is
\be
\bm{\Gamma}^{-1}=\frac1{\omega^2\mu}\bm{\nabla\times\nabla\times}
-\varepsilon,\label{dgminus1}
\ee
that is,
\be
\delta\bm{\Gamma}\cdot\bm{\Gamma}^{-1}=-\bm{\Gamma}\cdot
\delta\bm{\Gamma}^{-1}.
\ee
By comparing with the iterated source term in the vacuum-to-vacuum
persistence amplitude $\exp iW$, we see that an
infinitesimal variation of the bodies results in an effective source product,
\be
\mathbf{P(r)P(r')}\bigg|_{\rm eff}=i\delta\bm{\Gamma}^{-1},
\ee
from which we deduce from  Eq.~(\ref{w}) that 
\be
\delta W=\frac{i}2\mbox{Tr}\,\bm{\Gamma}\cdot\delta\bm{\Gamma}^{-1}=
-\frac{i}2\mbox{Tr}\,\delta\bm{\Gamma}\cdot\bm{\Gamma}^{-1}
=-\frac{i}2\delta \mbox{Tr}\,\ln\bm{\Gamma},\label{ddw}
\ee
where the trace includes integration over space-time coordinates.
We conclude, by ignoring an integration constant,
\be
W=-\frac{i}2\mbox{Tr}\,\ln\bm{\Gamma}.\label{trlog}
\ee
This is in precise analogy to the expression for scalar fields.
Another derivation of this result is given in the Appendix.
Incidentally, note that the first equality in Eq.~(\ref{ddw}) implies
for dielectric bodies ($\mu=1$)
\be
\delta W=-\frac{i}2\int\frac{d\omega}{2\pi}\int (d\mathbf{r})
\,\delta\varepsilon
(\mathbf{r},\omega)\Gamma_{kk}(\mathbf{r,r'};\omega),
\ee
which is the starting point for the derivation of the Lifshitz formula
\cite{lifshitz} in Ref.~\cite{Schwinger:1977pa}.

\section{Rederivation of Casimir-Polder formula}
Henceforth, let us consider pure dielectrics, that is, set $\mu=1$.
The free Green's dyadic, in the absence of
dielectric bodies,  satisfies the equation
\be
\left[\frac1{\omega^2}\bm{\nabla}\times\bm{\nabla}\times-1\right]\bm{\Gamma}_0
=\bm{1},
\ee
so the equation satisfied by the full Green's dyadic
is
\be
(\bm{\Gamma}_0^{-1}-V)\bm{\Gamma}=\mathbf{1},
\ee
where $V=\varepsilon-1$ within the body.  From this we deduce
immediately that
\be
\bm{\Gamma}=(1-\bm{\Gamma}_0V)^{-1}\bm{\Gamma}_0.
\ee
From the trace-log formula (\ref{trlog}) we see that the
energy for a static situation ($W=-\int dt\, E$) relative to
the free-space background is
\be
E=\frac{i}2\mbox{Tr}\,\ln\bm{\Gamma}_0^{-1}\cdot\bm{\Gamma}
=-\frac{i}2\mbox{Tr}\,\ln(1-\bm{\Gamma}_0V).
\ee
The trace here is only over spatial coordinates.
We will now consider the interaction between two
bodies, with non-overlapping potentials, $V=V_1+V_2$, where
$V_a=\varepsilon_a-1$ is confined to the interior of body $a$, $a=1,2$.
Although it is straightforward to proceed to write the
interaction between the bodies in terms of scattering operators,
for our limited purposes here, we will simply treat the potentials
as weak, and retain only the first, bilinear term in the interaction:
\be
E_{12}=\frac{i}2\mbox{Tr}\,\bm{\Gamma}_0V_1\bm{\Gamma}_0V_2.
\ee
Here, as may be verified by direct calculation \cite{levine},
\be
\bm{\Gamma}_0(\mathbf{r,r'})=\bm{\nabla}\times\bm{\nabla}\times \mathbf{1}G_0
(\mathbf{r-r'})-\mathbf{1}=(\bm{\nabla\nabla}-\mathbf{1}\zeta^2) G_0(
\mathbf{r-r'}),
\ee
where the scalar Helmholtz Green's function which satisfies causal or Feynman
boundary conditions is
\be
G_0(\mathbf{r-r'})=\frac{e^{-|\zeta|R}}{4\pi R},\quad R=|\mathbf{r-r'}|,
\ee
the Fourier transform of the Euclidean Green's function,
which obeys the differential equation
\be
(-\nabla^2+\zeta^2)G_0(\mathbf{r-r'})=\delta(\mathbf{r-r'}),
\ee
and $\zeta=-i\omega$.  

Thus the interaction between the two potentials is given by
\be
E_{12}=-\frac12\int\frac{d\zeta}{2\pi}\int (d\mathbf{r})(d\mathbf{r'})
\left[(\nabla_i\nabla_j-\zeta^2\delta_{ij})
\frac{e^{-|\zeta||\mathbf{r-r'}|}}{4\pi
|\mathbf{r-r'}|}\right]^2V_1(\mathbf{r})V_2(\mathbf{r'}).\label{e12}
\ee
The derivatives occurring here may be easily worked out:
\be
(\bm{\Gamma}_0)_{ij}=
(\nabla_i\nabla_j-\zeta^2\delta_{ij})\frac{e^{-|\zeta|R|}}{4\pi R}
=\left[-\delta_{ij}(1+|\zeta|R+\zeta^2R^2)+\frac{R_iR_j}{R^2}(3+3|\zeta|R|
+\zeta^2R^2)\right]\frac{e^{-|\zeta|R}}{4\pi R^3},
\ee
and then contracting two such factors together gives
\be(\nabla_i\nabla_j-\zeta^2\delta_{ij})\frac{e^{-|\zeta|R}}{4\pi R}
(\nabla_i\nabla_j-\zeta^2\delta_{ij})\frac{e^{-|\zeta|R}}{4\pi R}=
(6+12t+10t^2+4t^3+2t^4)\frac{e^{-2t}}{(4\pi R^3)^2},
\ee
where $t=|\zeta|R$.  Inserting this into Eq.~(\ref{e12}), we obtain
for the integral over $\zeta$
\be
-\frac1{64\pi^3R^7}\int_0^\infty du\,e^{-u}\left(6+6u+\frac52 u^2
+\frac12 u^3+\frac18 u^4\right)= -\frac{23}{64\pi^3 R^7},
\ee
or
\be
E_{12}=-\frac{23}{(4\pi)^3}\int (d\mathbf{r})(d\mathbf{r'})
\frac{V_1(\mathbf{r})V_2(\mathbf{r'})}{|\mathbf{r-r'}|^7},\label{cpv}
\ee
which is the famous Casimir-Polder potential \cite{cp}.
This formula is valid for bodies, which are presumed to be composed of 
material filling nonoverlapping volumes $v_1$ and $v_2$, respectively, 
characterized by
dielectric constants $\varepsilon_1$ and $\varepsilon_2$, both nearly unity.  
We emphasize that this formula is exact in the limit $\varepsilon_{1,2}\to1$,
as discussed in Ref.~\cite{Bordag:2001qi}.

\section{Energy of cylinder parallel to a plane}
In Ref.~\cite{Milton:2008vr} we derived the energy of two uniform dilute
cylinders, of radius $a$ and $b$ respectively, the parallel axes of which are
separated by a distance $R$, $R>a+b$.  In terms of the constant
\be
N=\frac{23}{640\pi^2}(\varepsilon_1-1)(\varepsilon_2-1).\label{N}
\ee
the energy of interaction per unit length is
\be 
\mathfrak{E}_{\rm cyl-cyl}=-\frac{32\pi N}3\frac{a^2b^2}
{R^6}\frac{1-\frac12\left(\frac{a^2+b^2}{R^2}\right)-\frac12
\left(\frac{a^2-b^2}{R^2}\right)^2}{\left[\left(1-\left(\frac{a+b}R\right)^2
\right)\left(1-\left(\frac{a-b}R\right)^2\right)\right]^{5/2}}.\label{cylcyl}
\ee If we take $R$ and $b$ to infinity, such that $Z=R-b$ is held fixed,
we describe a cylinder of radius $a$ parallel to a dielectric plane, 
where $Z$ is the distance between the axis of the cylinder and the 
plane.  That limit gives the simple result
\be
\mathfrak{E}_{\rm cyl-pl}=-\frac{N\pi a^2}{Z^4}\frac1{(1-a^2/Z^2)^{5/2}}.
\label{cyl-pl-E}
\ee
This is to be compared to the corresponding result for a sphere
of radius $a$ a distance $Z$ above a plane:
\be
E_{\rm sph-pl}=-N\frac{v}{Z^4}
\frac1{(1-a^2/Z^2)^2},\label{sp-pl-E}
\ee
where $v$ is the volume of the sphere.  

As an illustration of how the calculation is done, let us rederive
this result directly from Eq.~(\ref{cpv}). We see immediately that
the energy between an infinite halfspace (of permittivity $\varepsilon_1$)
 and a parallel slab (of permittivity $\varepsilon_2$) of area $A$
and thickness $dz$ separated by a distance $z$ is
\be
\frac{dE}{A}=-N\frac{dz}{z^4},
\ee
so the energy per length between the cylinder and the plane is
\be
\mathfrak{E}_{\rm cyl-pl}=
-2Na^2\int_{-1}^1 d\cos\theta\frac{\sin\theta}{(Z+a\cos\theta)^4}
=-\frac{N\pi a^2Z}{(Z^2-a^2)^{5/2}},
\ee
which is the result (\ref{cyl-pl-E}).

\section{Eccentric cylinders}

As a second illustration, consider two coaxial cylinders, of radii $a$ and
$b$, $a<b$. The inner cylinder is filled with material of permittivity
$\varepsilon_1$, while the outer cylinder is the inner boundary of 
a region with permittivity $\varepsilon_2$ extending out to infinity.
 An easy calculation from the van der Waals interaction 
(\ref{cpv})
\bea
\mathfrak{E}_{\rm co-cyl}&=&
-\frac{64 N}3\int_0^a d\rho\,\rho \int_b^\infty d\rho'\rho'
\int_0^{2\pi}\frac{d\theta}{(\rho^2+\rho^{\prime2}-2\rho\rho'\cos\theta)^3}
\nn\\
&=&-\frac{32 \pi N}{3}\int_0^{a^2} dx\int_{b^2}^\infty dy\frac{x^2+y^2+4xy}
{(y-x)^5}\nn\\
&=&-\frac{16 N\pi a^2b^2}{3(b^2-a^2)^3}.\label{cocyl}
\eea
This reduces to the dilute Lifshitz formula for the interaction between
parallel plates if we take the limit $b\to\infty$, $a\to\infty$, with
$b-a=d$ held fixed:
\be
\mathfrak{E}_{\rm co-cyl}\to -\frac{2N\pi b}{3d^3}, \quad\mbox{or}\quad
\frac{E}A=-\frac{N}{3d^3}.
\ee
Note that the result (\ref{cocyl}) may be obtained by analytically continuing
the energy between two externally separated cylinders, given by 
Eq.~(\ref{cylcyl}).  We simply take $R$ to zero there, and choose
the sign of the square root so that the energy is negative.  That
suggests that the same thing can be done to obtain  the energy
of interaction between two parallel cylinders, one inside the other, but
whose axes are displaced by an offset $R$, with $R+a<b$, as shown
in Fig.~\ref{fig1}:
\begin{figure}
\includegraphics{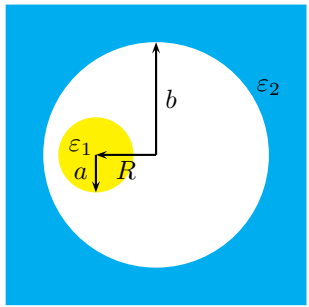}
\caption{\label{fig1} Dielectric $\varepsilon_2$ hollowed out by a
cylindrical cavity which contains an offset parallel dielectric
cylinder $\varepsilon_1$.}
\end{figure}
\be
\mathfrak{E}_{\rm ecc-cyl}
=-\frac{16\pi N}3\frac{a^2}{b^4}\frac{(1-a^2/b^2)^2+(1+a^2/b^2)
R^2/b^2-2R^4/b^4}{\left[(1-a^2/b^2)^2+R^4/b^4-2(1+a^2/b^2)R^2/b^2
\right]^{5/2}}.
\label{cc}
\ee
We can verify this is true by carrying out the integral from the 
Casimir-Polder formula (\ref{cpv})
\be
\mathfrak{E}_{\rm ecc-cyl}
=-\frac{32N}{3\pi}\int_0^a \rho\,d\rho\int_b^\infty \rho'\,
d\rho'\int_0^{2\pi}d\theta\int_0^{2\pi}d\theta'
\left[\rho^2+\rho^{\prime2}-2\rho\rho'\cos(\theta-\theta')+R^2
-2R(\rho\cos\theta-\rho'\cos\theta')\right]^{-3}.
\ee
The angular integrals can be done, but the remaining integrals are rather
complicated. Therefore, let us simply expand immediately in the quantities
$x=\rho/\rho'$ and $y=R/\rho'$, which are both less than one.  Then we
can carry out the four integrals term by term.  In this way we
find
\bea
\mathfrak{E}_{\rm ecc-cyl}&=&-\frac{16\pi Na^2}{3b^4}
\sum_{n,m=0}^\infty \left(\frac{a^2}{b^2}\right)^n\left(\frac{R^2}{b^2}
\right)^m\frac{(m+1)^2}{2}{n+m+1\choose m+1}{n+m+2\choose m+1},
\eea
which is exactly the series expansion of Eq.~(\ref{cc}) for small $a/b$
and $R/b$.

By differentiating this energy with respect to the offset $R$, we obtain
the force of interaction between the inner cylinder and the outer one,
$\mathfrak{F}=-\partial\mathfrak{E}/\partial R$.  Evidently, that force
is zero for coaxial cylinders, since that is a point of unstable
equilibrium.  For small $R$, $\mathfrak{F}$ grows  linearly with $R$
with a positive coefficient.  The inner cylinder is attracted to the 
closest point of the opposite cylinder.  Similar considerations
for conducting cylinders were given in 
Refs.~\cite{Mazzitelli:2006ne,Dalvit:2006wy}, 
with the idea that the cylindrical geometry might
prove to be a useful proving ground for Casimir experiments.

\section{Self-energy of dilute cylinder}

This is a rederivation of the result found by dimensional continuation in 
Ref.~\cite{Nesterenko:1997fq}.
The summation of the Casimir-Polder forces between the molecules in
a single cylinder of radius $a$ 
is given by (in $N$ $\varepsilon_1=\varepsilon_2$)
\bea
\mathfrak{E}_{\rm cyl}&=&-\frac{32N}3\int_0^a d\rho\,\rho
\int_0^a d\rho'\,\rho'\int_0^{2\pi}\frac{d\theta}{(\rho^2+\rho^{\prime2}
-2\rho\rho'\cos\theta)^3}\nn\\
&=&-\frac{16N}3\int_0^{a^2}\frac{dx}{x^2}\int_0^1 du\left(\frac1{u^3}
-\frac6{u^4}+\frac6{u^5}\right).
\eea
This is, of course, terribly divergent.  We can regulate it by
analytic continuation: replace the highest power of 
$(\rho^2-\rho^{\prime2})^{-1}=(xu)^{-1}$, 5, by $\beta$,
and regard $\beta$ as less than $1$.  Then,
\bea
\mathfrak{E}_{\rm cyl}&=&-\frac{16N}3\int_0^{a^2} dx \, x^{3-\beta}
\int_0^1du\left(u^{2-\beta}-6u^{1-\beta}+6u^{-\beta}\right)\nn\\
&=&-\frac{16N}3(a^2)^{4-\beta}\frac{(5-\beta)}{(1-\beta)(2-\beta)
(3-\beta)}.
\eea
Now if we analytically continue to $\beta=5$ we get an vanishing
self-energy to order $(\varepsilon-1)^2$.  This result was first
discovered by Romeo (private communication), verified in 
Ref.~\cite{Nesterenko:1997fq}, and confirmed later by full
Casimir calculations \cite{CaveroPelaez:2004xp,Romeo:2005qk}.

\begin{acknowledgments}
We thank the US National Science Foundation (Grant No.\ PHY-0554926) and the
US Department of Energy (Grant No.\ DE-FG02-04ER41305) for partially funding
this research.  We thank Archana Anandakrishnan and In\'es Cavero-Pel\'aez 
for extensive collaborative assistance throughout this project,
and Simen \AA. Ellingsen for bringing Ref.~\cite{Franchek} to our attention.
Particularly, we wish to express our gratitude to Iver Brevik for
sharing his deep understanding of electromagnetic theory with us
over many years.

\end{acknowledgments}

\appendix
\section{Derivation of Green's dyadic formalism from canonical
theory}
Here we begin by sketching the development of the Green's dyadic
equation from  canonical quantum electrodynamics.  For simplicity
of the discussion, we will consider a medium without dispersion, so
that $\varepsilon$ and $\mu$ are constant.  First we must state
the canonical equal-time commutation relations.  
We will require (only transverse fields are relevant)
\be
[E_i(\mathbf{r},t),E_j(\mathbf{r'},t)]=0.
\ee
In a medium, it is the electric displacement field which is canonically
conjugate to the vector potential, so we have the equal-time commutation
relation (Coulomb gauge)
\be
[\mathbf{A}(\mathbf{r},t),\partial_0 \mathbf{A}(\mathbf{r'},t)]
=\frac{i}\varepsilon\left(\mathbf{1}
-\frac{\bm{\nabla\nabla}}{\nabla^2}\right)\delta(\mathbf{r-r'}).\label{etcr}
\ee
Now in view of Eq.~(\ref{vev}), and the Maxwell equations
\begin{subequations}
\bea
\bm{\nabla}\times \mathbf{E}&=&-\frac\partial{\partial t}\mathbf{B},\label{me1}
\\
\bm{\nabla}\times \frac1\mu\mathbf{B}&=&\frac\partial{\partial t}
\varepsilon\mathbf{E},
\eea
\end{subequations}
we deduce from Eq.~(\ref{vev})
\be
\bm{\nabla\times\nabla\times\Gamma'}+\varepsilon\mu\frac{\partial^2}{\partial
t^2}\bm{\Gamma'}=i\varepsilon\mu\delta(t-t')\langle[\dot {\mathbf{E}}
(\mathbf{r},t), \mathbf{E}(\mathbf{r'},t)]\rangle.\label{gf1}
\ee
But according to Maxwell's equations and Eq.~(\ref{etcr})
\be
[\dot{\mathbf{E}}(\mathbf{r},t),\mathbf{E(r'},t)]
=\frac1{\varepsilon\mu}\bm{\nabla\times\nabla\times}[\mathbf{A(r},t),
-\partial_0\mathbf{A(r'},t)]=-\frac{i}{\varepsilon^2\mu}\bm{\nabla\times
\nabla\times 1}\delta(\mathbf{r-r'}).
\ee
If we now insert this into Eq.~(\ref{gf1}) we obtain for the
Fourier transform of the Green's dyadic (\ref{ftgd})
\be
(\bm{\nabla\times\nabla\times}
-\varepsilon\mu\omega^2)\bm{\Gamma'}=\frac1\varepsilon
\bm{\nabla\times\nabla\times 1},
\ee
which is indeed the equation satisfied by the solenoidal Green's
dyadic, Eq.~(\ref{gprimeeq}).

It is equally easy to derive the trace-log formula.  We have the
variational statement, for infinitesimal changes in the permittivity
and the permeability \cite{embook},
\be
\delta E=-\frac12\int (d\mathbf{r})\langle \delta \varepsilon E^2+\delta\mu H^2
\rangle.
\ee
Given Eqs.~(\ref{vev}) and (\ref{me1}) we can write this in terms of the
coincident-point limit of Green's dyadic,
\be
\delta E=\frac{i}2\int (d\mathbf{r})\left[\delta\varepsilon
-\delta\left(\frac1\mu\right)\frac1{\omega^2}\bm{\nabla\times\nabla\times}
\right]\bm{\Gamma}(\mathbf{r,r'})\bigg|_{\mathbf{r'\to r}}=-\frac{i}2
\mbox{Tr}\,\delta\bm{\Gamma}^{-1}\cdot\bm{\Gamma},
\ee
according to Eq.~(\ref{dgminus1}),
which involves an integration by parts, and coincides with 
the first equality in Eq.~(\ref{ddw}).


\begin{thebibliography}{99}


\bibitem{Casimir:1948dh}
  H.~B.~G.~Casimir,
 Kon.\ Ned.\ Akad.\ Wetensch.\ Proc.\  {\bf 51}, 793 (1948).


\bibitem{Wirzba:2007bv}
  A.~Wirzba,
  J.\ Phys.\ A  {\bf 41}, 164003 (2008).
[arXiv:0711.2395 [quant-ph]].


 

\bibitem{Bordag:2008gj}
  M.~Bordag and V.~Nikolaev,
  J.\ Phys.\ A  {\bf 41}, 164002 (2008).
  [arXiv:0802.3633 [hep-th]].




\bibitem{maianeto08}
P. A. Maia Neto, A. Lambrecht, S. Reynaud, Phys.\ Rev.\ A {\bf 78}, 012115
(2008) [arXiv:0803.2444].
 

\bibitem{Emig:2007cf}
 T.~Emig, N.~Graham, R.~L.~Jaffe and M.~Kardar,
 Phys.\ Rev.\ Lett.\   {\bf99}, 170403 (2007)
[arXiv:0707.1862 [cond-mat.stat-mech]];
  Phys.\ Rev.\  D {\bf 77}, 025005 (2008)
[arXiv:0710.3084 [cond-mat.stat-mech]];
  T.~Emig and R.~L.~Jaffe,
  J.\ Phys.\ A  {\bf 41}, 164001 (2008)
[arXiv:0710.5104 [quant-ph]].

\bibitem{pft} J. Blocki, J. Randrup, W. J. \'Swi\c{a}tecki, and C. F.
Tsang, Ann.\ Phys.\ (N.Y.) {\bf 105}, 427 (1977).

\bibitem{derjaguin} B. V. Deryagin, Kolloid Z. {\bf 69}, 155 (1934).

\bibitem{Onofrio:2006mq}
  R.~Onofrio,
  New J.\ Phys.\  {\bf 8}, 237 (2006)
  [arXiv:hep-ph/0612234].

\bibitem{Gies:2006cq}
  H.~Gies and K.~Klingm\"uller,
  Phys.\ Rev.\  D {\bf 74}, 045002 (2006)
  [arXiv:quant-ph/0605141].

\bibitem{Gies:2006bt}
  H.~Gies and K.~Klingm\"uller,
  Phys.\ Rev.\ Lett.\  {\bf 96}, 220401 (2006)
  [arXiv:quant-ph/0601094].


\bibitem{Gies:2006xe}
  H.~Gies and K.~Klingm\"uller,
  Phys.\ Rev.\ Lett.\  {\bf 97}, 220405 (2006)
  [arXiv:quant-ph/0606235].


\bibitem{Gies:2003cv}
  H.~Gies, K.~Langfeld and L.~Moyaerts,
  JHEP {\bf 0306}, 018 (2003)
  [arXiv:hep-th/0303264].















\bibitem{Scardicchio:2004fy}
  A.~Scardicchio and R.~L.~Jaffe,
  Nucl.\ Phys.\  B {\bf 704}, 552 (2005)
 [arXiv:quant-ph/0406041].


\bibitem{Scardicchio:2005di}
  A.~Scardicchio and R.~L.~Jaffe,
  Nucl.\ Phys.\  B {\bf 743}, 249 (2006)
  [arXiv:quant-ph/0507042].



\bibitem{Fulling03}
S.~A.~Fulling,
Proc.~6th Workshop on Quantum Field Theory Under the Influence of
External Conditions, ed.~K.~A.~Milton (Rinton Press, Princeton, NJ,
2004), p.~166.

\bibitem{fulling08} S. A. Fulling, L. Kaplan, K. Kirsten, Z. H. Liu,
and K. A. Milton, arXiv:0806.2468.



\bibitem{capasso}
A. Rodriguez, M. Ibanescu, D. Iannuzzi, F. Capasso, J. D. Joannopoulos,
and S. G. Johnson, 
Phys.\ Rev.\ Lett.\ {\bf99}, 080401 (2007) [arXiv:0704.1890v2].

\bibitem{rodriguez}
A. Rodriguez, M. Ibanescu, D. Iannuzzi, J. D. Joannopoulos, and S. G. Johnson, 
Phys.\ Rev.\ A {\bf76},032106 (2007) 

\bibitem{rod} A. W. Rodriguez, J. D. Joannopoulos, and S. G. Johnson, 
Phys.\ Rev.\ A {\bf 77}, 062107 (2008).

\bibitem{krein} M. G. Krein, Mat.\ Sb. (N.S.) {\bf33}, 597 (1953);
Dokl.\ Akad.\ Nauk SSSR {\bf144}, 475 (1962) [Sov.\ Math.\ Dokl. {\bf3},
707 (1962)]; M. Sh. Birman and M. G. Krein, Dokl.\ Akad.\ Nauk SSSR {\bf144},
268 (1962) [Sov.\ Math.\ Dokl.\ {\bf 3}, 740 (1962)].

  
\bibitem{renne} M. J. Renne, Physica {\bf 56}, 125 (1971).

\bibitem{balian}
  R.~Balian and B.~Duplantier,
  arXiv:quant-ph/0408124, 
  in the proceedings of 15th SIGRAV Conference on General Relativity and 
  Gravitational Physics, Rome, Italy, 9--12 September 2002;
  Ann.\ Phys.\ (N.Y.)  {\bf 112}, 165 (1978);
  Ann.\ Phys.\ (N.Y.)  {\bf 104}, 300 (1977).


\bibitem{Milton:2007wz}
  K.~A.~Milton and J.~Wagner,
  J.\ Phys.\ A  {\bf 41}, 155402 (2008)
  [arXiv:0712.3811 [hep-th]].





\bibitem{Milton:2007gy}
  K.~A.~Milton and J.~Wagner,
  Phys.\ Rev.\  D {\bf 77}, 045005 (2008)
  [arXiv:0711.0774 [hep-th]].

\bibitem{CaveroPelaez:2008tj}
 I.~Cavero-Pel\'aez, K.~A.~Milton, P.~Parashar and K.~V.~Shajesh,
Phys.\ Rev.\ D {\bf 78}, 065018, 065019 (2008)
[arXiv:0805.2776 [hep-th],  
 arXiv:0805.2777 [hep-th]];
J. Wagner, K.~A.~Milton, and P.~Parashar, submitted to 60 Years of the
Casimir Effect, Brasilia, June 2008,  arXiv:0811.2442 [hep-th].


\bibitem{Bordag:2008rc}
  M.~Bordag and N.~Khusnutdinov,
  Phys.\ Rev.\  D {\bf 77}, 085026 (2008)
  [arXiv:0801.2062 [hep-th]].

\bibitem{Bordag:2007zz}
  M.~Bordag,
  Phys.\ Rev.\  D {\bf 76}, 065011 (2007)
[arXiv:0704.3845].

\bibitem{Bordag:2006kx}
  M.~Bordag,
  Phys.\ Rev.\  D {\bf 75}, 065003 (2007)
  [arXiv:quant-ph/0611243].



\bibitem{Milton:2008vr}
  K.~A.~Milton, P.~Parashar and J.~Wagner,
Phys. Rev. Lett. {\bf 101}, 160402 (2008)  [arXiv:0806.2880 [hep-th]].



\bibitem{Milton:1997ky}
  K.~A.~Milton and Y.~J.~Ng,
  Phys.\ Rev.\  E {\bf 57}, 5504 (1998)
  [arXiv:hep-th/9707122].


\bibitem{Nesterenko:1997fq}
K.~A.~Milton,  A.~V.~Nesterenko and V.~V.~Nesterenko,
  Phys.\ Rev.\  D {\bf 59}, 105009 (1999)
  [arXiv:hep-th/9711168v3].


\bibitem{Bordag:2001qi}
  M.~Bordag, U.~Mohideen and V.~M.~Mostepanenko,
  Phys.\ Rept.\  {\bf 353}, 1 (2001)
  [arXiv:quant-ph/0106045].

\bibitem{Franchek}
S. W. Montgomery, M. A. Franchek, and V. W. Goldschmidt,
J. Coll.\ Interface Sci.\ {\bf227}, 567 (2000). 

\bibitem{Schwinger:1977pa}
  J.~Schwinger, L.~L.~DeRaad, Jr., and K.~A.~Milton,
  Ann.\ Phys.\ (N.Y.)  {\bf 115}, 1 (1979).

\bibitem{lifshitz} E. M. Lifshitz, Zh. Eksp. Teor. Fiz. {\bf 29}, 94 (1956)
[Sov.\ Phys.\ JETP {\bf2}, 73 (1956)].

\bibitem{levine}
H. Levine and J. Schwinger, Comm. Pure Appl.\ Math.\ III {\bf 4}, 
355 (1950).

\bibitem{cp}
H. B. G. Casimir and D. Polder, Phys.\ Rev.\ {\bf 73}, 360 (1948).


\bibitem{Mazzitelli:2006ne}
  F.~D.~Mazzitelli, D.~A.~R.~Dalvit and F.~C.~Lombardo,
  New J.\ Phys.\  {\bf 8}, 240 (2006)
  [arXiv:quant-ph/0610181].


\bibitem{Dalvit:2006wy}
  D.~A.~R.~Dalvit, F.~C.~Lombardo, F.~D.~Mazzitelli and R.~Onofrio,
  Phys.\ Rev.\  A {\bf 74}, 020101 (2006).



\bibitem{CaveroPelaez:2004xp}
  I.~Cavero-Pel\'aez and K.~A.~Milton,
  Ann.\ Phys.\ (N.Y.) {\bf 320}, 108 (2005)
  [arXiv:hep-th/0412135].


\bibitem{Romeo:2005qk}
  A.~Romeo and K.~A.~Milton,
  Phys.\ Lett.\  B {\bf 621}, 309 (2005)
  [arXiv:hep-th/0504207].




\bibitem{embook}
J. Schwinger, L. L. DeRaad, Jr., K. A. Milton, and W.-y. Tsai,
{\it Classical Electrodynamics} (Perseus/Westview, New York, 1998). 










\end{thebibliography}
\end{document}